# Potential of RPCs for tracking

T. Francke[1], P. Fonte[2], V. Peskov[1,*], J. Rantanen[3]

*1-Royal Institute of Technology, Stockholm, Sweden,*

*2 - LIP and ISEC, Coimbra, Portugal,*
*3 - XCounter AB, Danderyd, Sweden.*

## Abstract

We have demonstrated that small gap (0.1 to 0.4 mm) RPCs made of low resistivity materials ($\rho<10^8\ \Omega\cdot$cm) can operate at counting rates of up to $10^5$ Hz/mm$^2$ with position resolutions better than 50 µm. Results of preliminary tests allow us to suggest a possible application of this new type of RPC for tracking.

* Corresponding author: V. Peskov, Physics Department, Royal Institute of Technology, Frescativagen 24, Stockholm 10405, Sweden (telephone: 468161000, e-mail: vladimir.peskov @cern.ch).

## I. Introduction

In a previous work [1] we have demonstrated that a very good position resolution (better than 50 μm) may be achieved for X-ray detection (6-30 keV) using a small gap (below 0.5mm) RPC with the cathode covered with a CsI secondary electron emitter. In a subsequent work [2] we succeeded to considerably improve the counting rate characteristics of the RPC by using low resistive materials for its electrodes. In this work we aims to:

1) combine both approaches and develop a high rate and position resolution RPC;

2) investigate the possibility of applying this type of RPC for tracking in High-Energy Physics experiments.

## II. Experimental set up

The experimental set up (Fig.1) consisted of an aluminium gas container inside which various RPC designs (with sizes $5\times5cm^2$ and $10\times10cm^2$) could be installed and tested. To minimize the voltage drop across the resistive plates and achieve good rate characteristics the cathodes were manufactured from low resistivity materials, for example Pestov glass ($\rho\sim10^9$ Ω·cm) [3], conductive plastics ($\rho=10^6-10^8$ Ω·cm) [4], GaAs ($\rho=10^{-1}-10^8$ Ω·cm) [5] and Si ($\rho=10^{-2}-10^4$ Ω·cm) [5]. The anode electrode was manufactured from Pestov glass ($\rho\sim10^{10}$ Ωcm). The inner (gas side) faces of the anode electrodes had Chromium readout strips deposited at either 30 or 50 μm pitch, allowing highly accurate position measurements.

Two main designs were used. The first design had 20 or 30 readout strips in the central region of the anode and the rest of the anode was metalised. The second design had strips over the whole area, being the strips in the central region connected to individual amplifiers whereas the others were connected in groups. The gap between electrodes could be varied from 0.1 to 0.5 mm.

Before assembling the RPCs they were first ultrasonically cleaned in a solution of dish soap and distilled water and then ultrasonically cleaned several times in distilled deionised water to remove dust and any other microparticles attached to the surfaces. The test chamber was also carefully cleaned in the same way. Gas pipes were equipped with 3 μm microparticle filters. Tests were performed in Ar, Kr and Xe-based gas mixtures with various quenchers ($CO_2$, ethane, isobutane, alcohol and tetrafluoroethane[†]) and their combinations. The percentages of each component were widely varied.

Ionisation tracks inside the RPC were produced by cosmic muons, β-particles and by high energy (>20keV) X-ray generated photoelectrons. Cosmic muons were identified by coincidences with the signals from two scintillators (see Fig.1).

Some position resolution measurements were performed with β-particles from $^{106}$Ru or $^{90}$Sr sources. The particles penetrated to the RPC through a 100 μm lead slit placed on the outer cathode surface (see details in Fig2). The cathode region around the slit had a thickness of 10 μm only and its inner part was covered with a porous CsI secondary electron emitter, 5 to 10 μm thick for increasing the number of primary electrons created near the cathode [6]. However, porous CsI emitters are easily charged up and can therefore be only used at low rates, typically <$10^3$Hz/cm$^2$.

[†] $C_2H_2F_4$, commercially known as R134a.

Position resolution measurements were also made with X-ray generated photoelectron tracks. The incoming photons were collimated by a 30 μm wide slit oriented perpendicularly to the electrode planes, which could be moved perpendicularly to the strips with an accuracy of a few μm. The slit was illuminated by an X-ray gun able to produce in excess of $10^5$ counts/mm$^2$ in the detector volume. In most measurements the cathode was covered with a uniform layer (0.3 - 0.4 μm thick) of CsI [1,6], which does not exhibit any strong charging up effect. If X-ray photons enter the CsI emitter at a shallow angle, there will be a sizable absorption probability, while at the same time secondary electrons created after the X-ray absorption can be easily emitted in to the detector volume [1]. Counters with cathodes not coated with the CsI layer were tested as well.

Avalanche-generated signals from the RPCs were measured at low rates with charge sensitive amplifiers, while current amplifiers were used at high rates and for discharge studies.

## III. Results

### III.1 Gain-rate characteristics of RPCs made of GaAs and Si

As was shown in [2], the use of low resistivity materials for the RPC electrodes allows one to reach a high counting rate and simultaneously ensure the protection of the electronics from damage in case of occasional discharges. In this work we have mostly used GaAs and Si, which were first reported for RPCs in [5]. In Fig.3 gain vs. rate curves for low resistivity RPCs made of GaAs and Si are presented, as well as the maximum achievable gain for the metallic parallel-plate avalanche chamber (PPAC). One can see that at gains above $10^4$, counting rates can be achieved up to $10^5$ Hz/mm$^2$. In principle one can run the counters at much higher gains but at the expense of a lower achievable rate (Fig. 3). In any case, the GaAs and Si-based RPCs both reach the, in principle insurmountable, rate limits of a metallic PPAC. However it should be noted that the rate limits for PPACs and RPCs might be set by different mechanisms (see next section and [7, 8] for more details).

### III-2. Discharges and their suppression in high-rate RPCs

As was shown in our previous studies, the maximum achievable gain in PPACs with metallic electrodes drops with increasing counting rate due to spark discharges [8]. It was found that the same is true for RPCs with low resistivity electrodes (<$10^4$ Ω·cm). However, the breakdown characteristics of a “medium” resistivity RPC is very different from the metallic PPAC or the high resistivity RPC. In the resistivity range $10^7$-$10^8$ Ω·cm, a new phenomenon - a continuous glow discharge – appears [2]. From the point of view of possible electronics damage, this type of discharge could be in some cases even more dangerous than the sparks. To identify the conditions at which sparks or glow discharges occur we performed studies with a wide range of electrode resistivities in various gas mixtures with a quencher concentration lower than 20%. The results of such measurements are summarized in Fig 4.

It was also found that the duration of the glow discharge can be decreased to a fraction of μs or even fully suppressed by using highly quenched mixtures, for example a high (20-25%) concentration of ethane in argon. Note for comparison that in the sparking region highly quenched mixtures do not have any effect [9].

### III.3 Position resolution

#### III-3a) Results with cosmic rays

As an example, Fig.5 shows signals from three adjacent anode strips obtained in coincidence with signals from two scintillators. In ~30% of the cases the large signal appears mainly on one of the strips, in 40% on two strips and in 30% of the cases in more than two strips. These results suggest that a rather good position resolution with minimum ionising particles should be achievable.

#### III-3b) Results with *b* particles

Fig. 6 shows the induced charge profile for *b* particles penetrating to the RPC through a 100 μm slit, with a FWHM of ~300 μm. A constant level background was subtracted from the points and the peak value normalized.)

One should note that the *b* particles strongly scatter in the collimator and cathode material and thus the effective collimation width for the penetrating particles could be larger than 100 μm.

#### III-3c) Results obtained with photoelectron tracks at rates up to $10^5$Hz/mm$^2$.

Photoelectrons generated by energetic X-rays have a mean free path in Ar, Kr, Xe at atmospheric pressure up to 2 or 3 mm [10, 11] and thus can be used to simulate charged-particle tracks. The main advantages of this method are that X-rays are much easier to collimate than *b* particles and that intense artificial sources are available, allowing the use of narrow slits.

Fig. 6 shows the induced charge profile when a well-collimated X-ray beam (30 μm width) was introduced to the amplification gap parallel to the cathode, approximately 50 μm apart from its surface and the beam oriented along the anode strips. The profile width is ~200 μm FWHM.

Fig. 7a shows the number of counts from various strips (a digital image) for the same irradiation conditions described in the previous paragraph. Fig. 7b shows the image of the same slit shifted by 25 microns in a direction perpendicular to the strips. The pattern is periodically repeated when the beam is further moved across the strips. From the image contrast (ratio of counts from neighbouring strips) one can conclude that a position resolution better than 30 μm was achieved.

Fig. 7c shows the image of a 7 lp/mm phantom: three slits 70 μm apart were easily resolved.

Note that due to the anti-parallax feature of the parallel plate gas amplification structure, in which the maximum multiplication is obtained only for the electrons created near the cathode, almost the same position resolution was achieved without a CsI converter as in Fig. 7 a) and b).

## IV. Conclusions and Outlook

We have demonstrated that a small gap RPC made of low resistive materials combines a high counting rate capability (up to $10^5$ Hz/mm$^2$ at gains up to $10^5$), approaching that of metallic PPACs, with an extremely good position resolution (aprox.30 μm) and still providing efficient electronics protection. This excellent position accuracy was obtained in a simple counting (digital) mode without using any analogue interpolation method. Also note that small gap RPCs enjoy extremely good timing properties (~50 ps *s* [12]). All these characteristics make small gap RPCs potentially competitive to other detectors in many applications, for example in medical imaging [13], biology, crystallography and also possibly as a high-rate tracking device.

## Figure captions

Fig. 1. General schematic drawing of the experimental set-up.

Fig. 2. Schematic drawing of the *b*-particles collimation setup.

Fig. 3. Gain vs. rate for RPCs made of various low resistivity materials: 1) $\rho=3\times10^{8}\Omega\cdot cm^{2}$); 2) $\rho= 4\times10^{7}\Omega\cdot cm^{3}$); 3) $\rho\sim10\times4\Omega\cdot cm$ (GaAs and Si); 4) gain vs rate for metallic PPAC.

Fig. 4. Types of discharges in RPCs with various electrode resistivities.

Fig. 5. Signals from three adjusted strips in the case of detection of the cosmic event. Horizontal scale 0.5ms/div; vertical scale 0.5V/div.

Fig. 6. Induced charge profile for the detection of collimated *b* particles and X-rays.

Fig. 7. Number of counts from various strips in the case when X-rays penetrated to the RPC through a 30 μm slit with the beam oriented along the strip #8 (a) and in between the strips #8 and #9 (b). The picture was periodically repeated when the beam was further moved perpendicularly to the strips. c) Number of counts from various strips for X-ray imaging of a 7 lp/mm phantom.

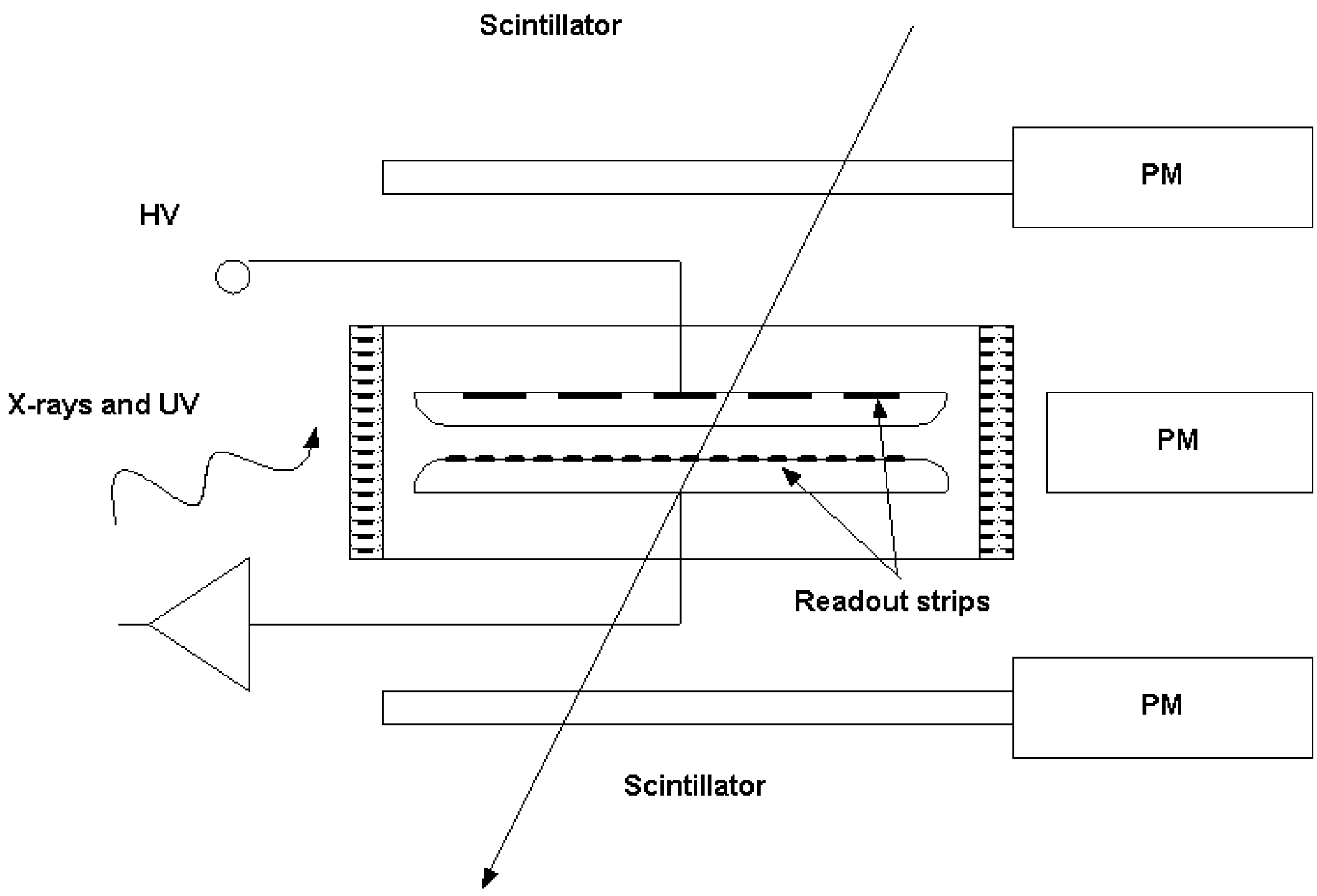

Fig. 1. General schematic drawing of the experimental set-up.

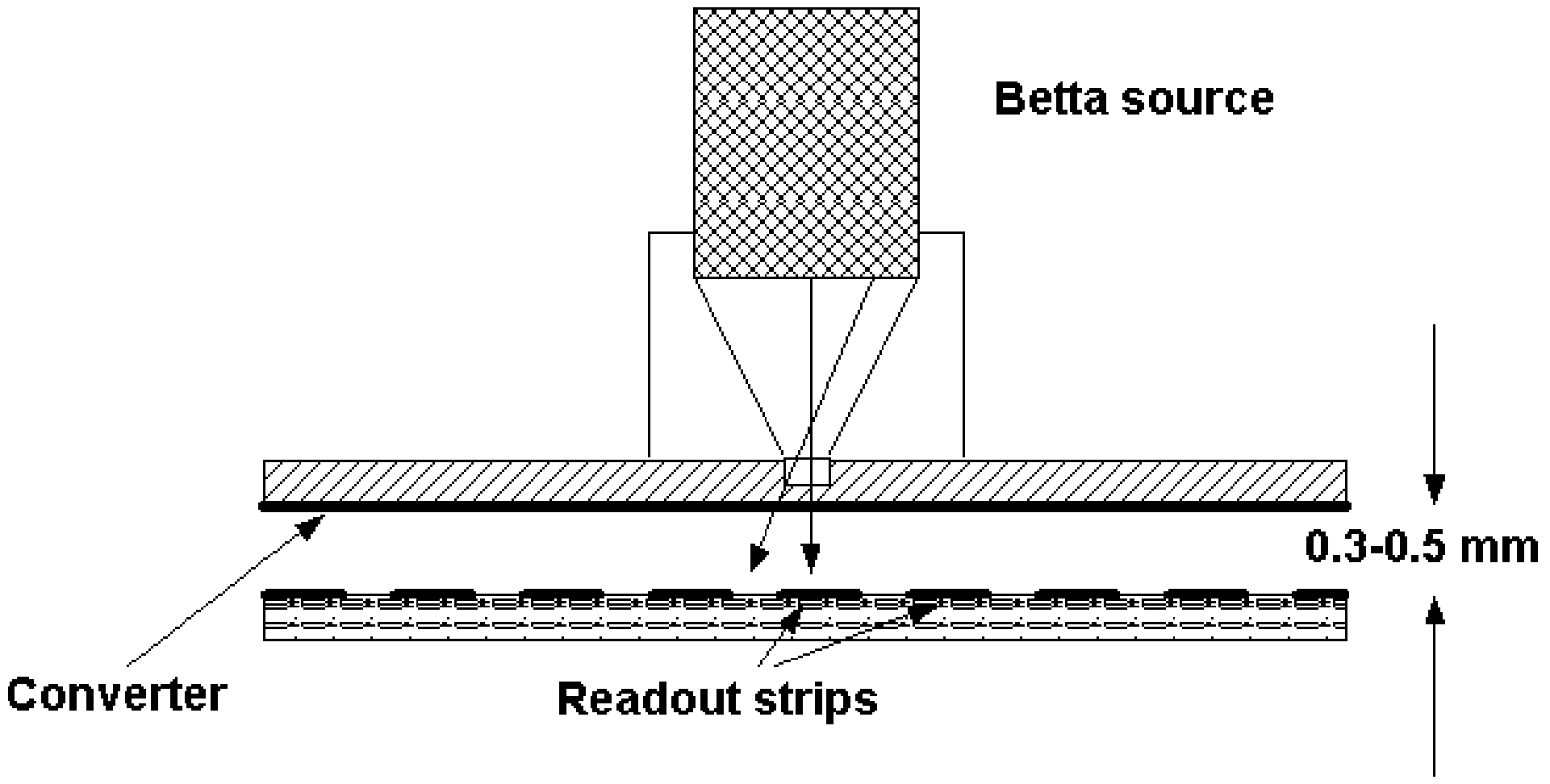

Fig. 2. Schematic drawing of the β-particles collimation setup.

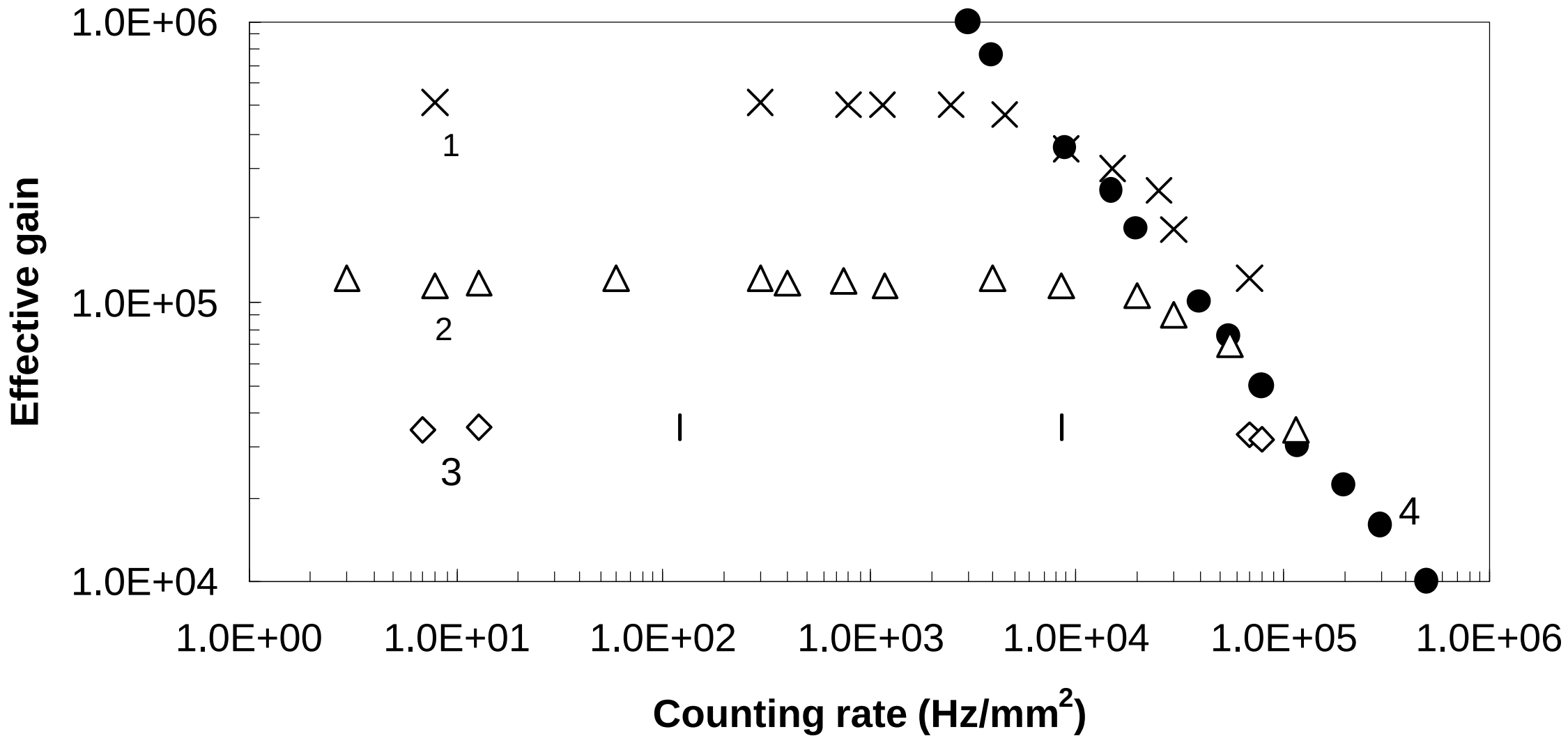

Fig. 3. Gain vs. rate for RPCs made of various low resistivity materials: 1) ρ=3×108Ω·cm2); 2) ρ= 4×107Ω·cm3); 3) ρ~10×4Ω·cm (GaAs and Si); 4) gain vs rate for metallic PPAC.

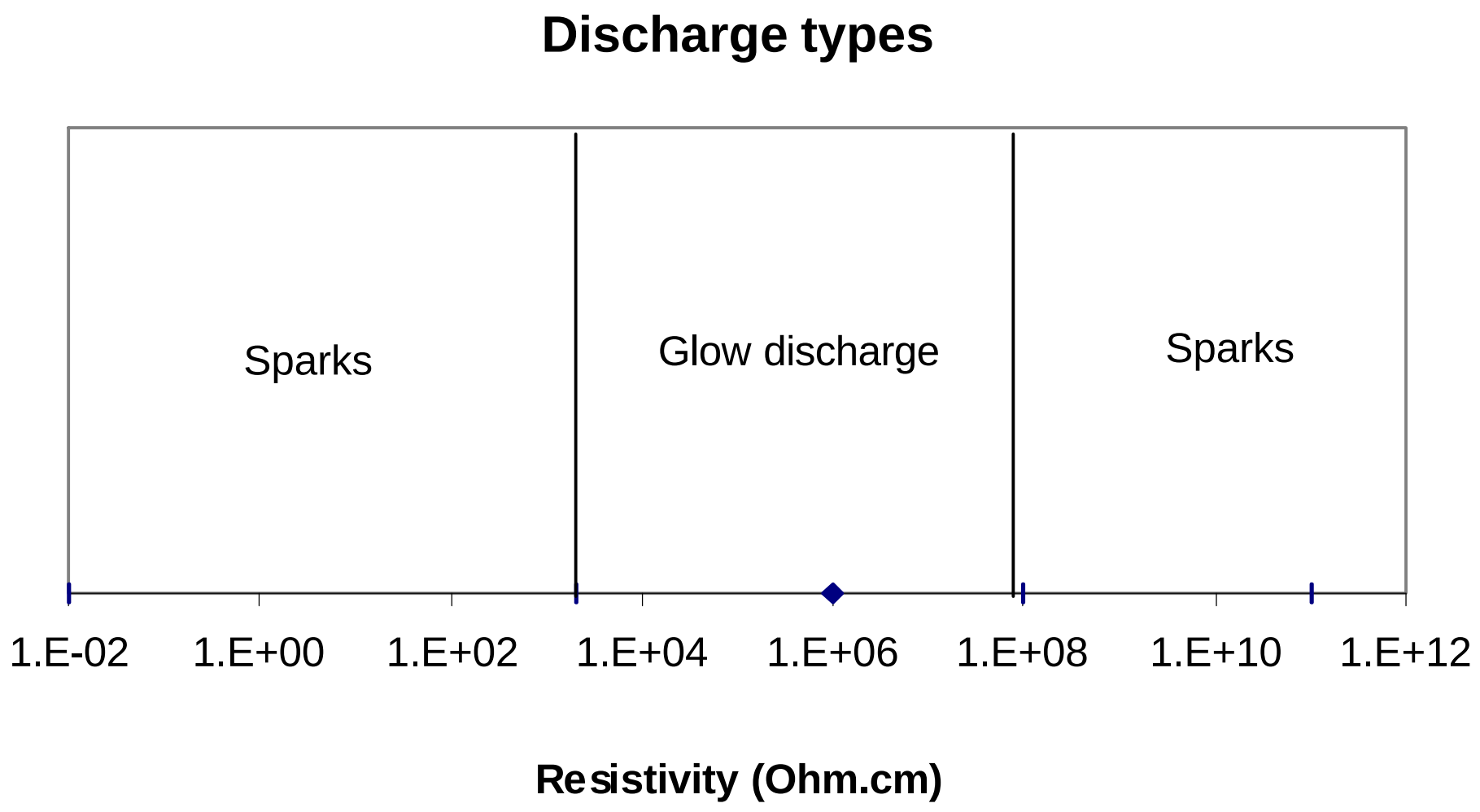

Fig. 4. Types of discharges in RPCs with various electrode resistivities.

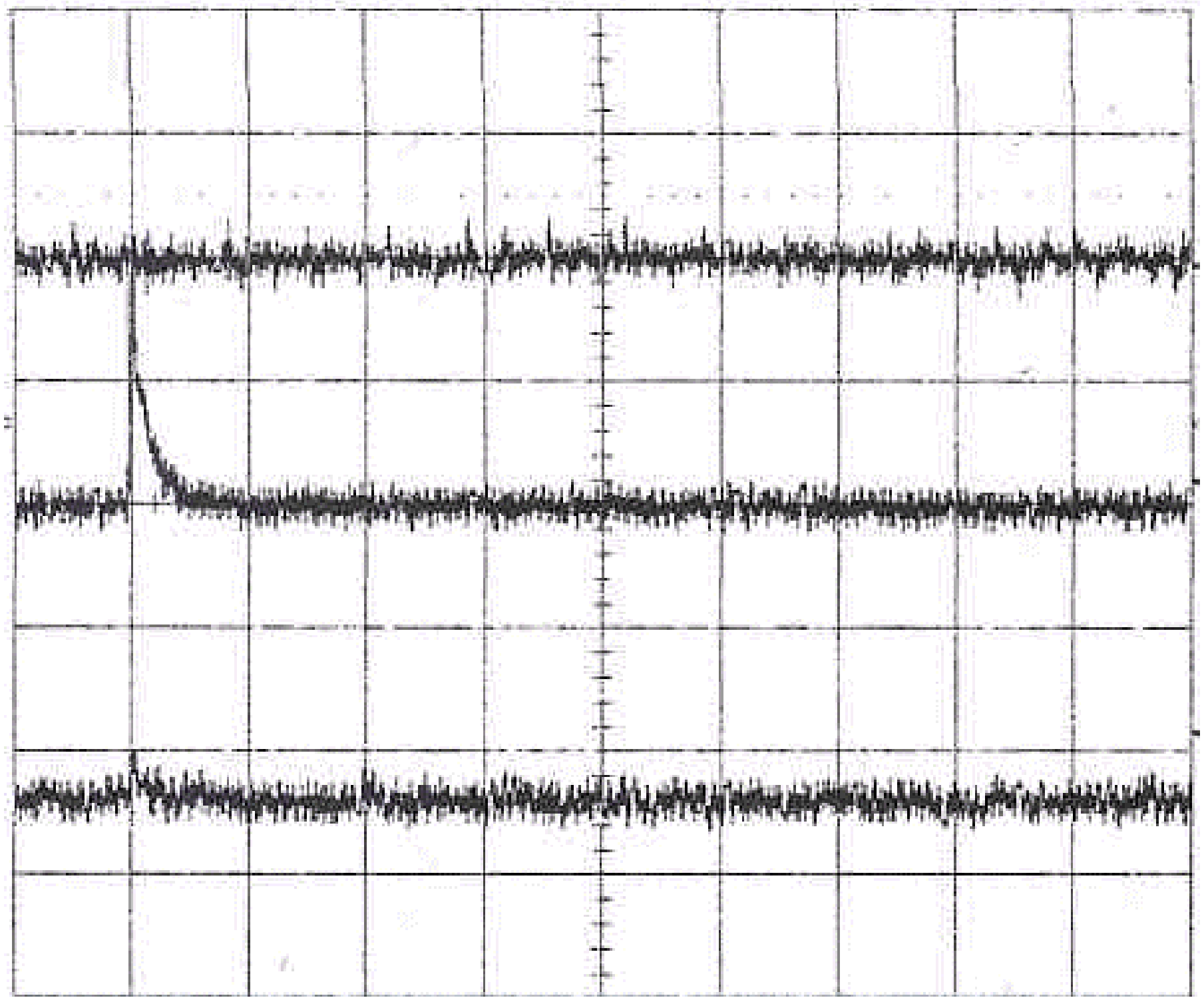

Fig. 5. Signals from three adjusted strips in the case of detection of the cosmic event. Horizontal scale 0.5ms/div; vertical scale 0.5V/div.

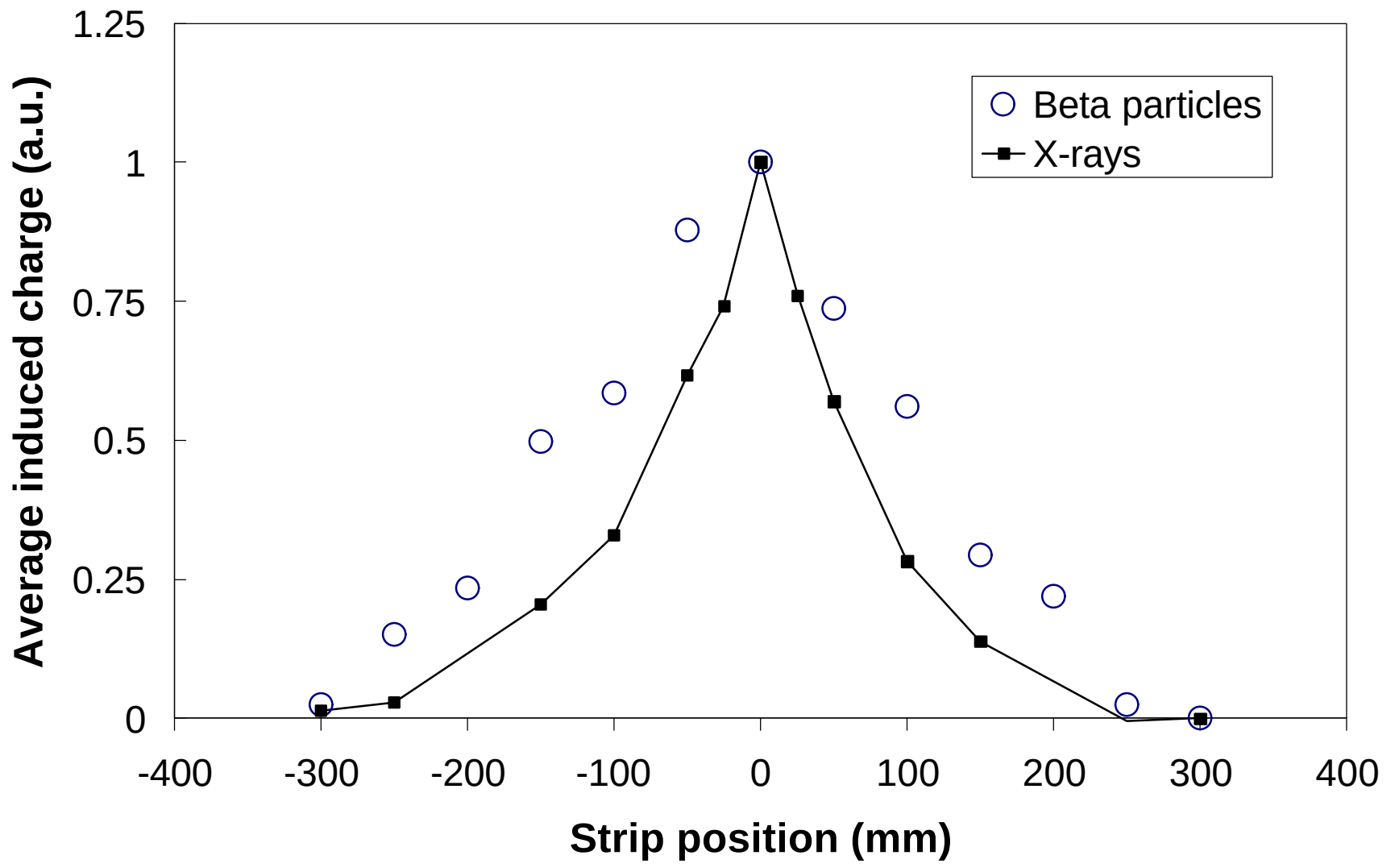

Fig. 6. Induced charge profile for the detection of collimated β particles and X-rays.

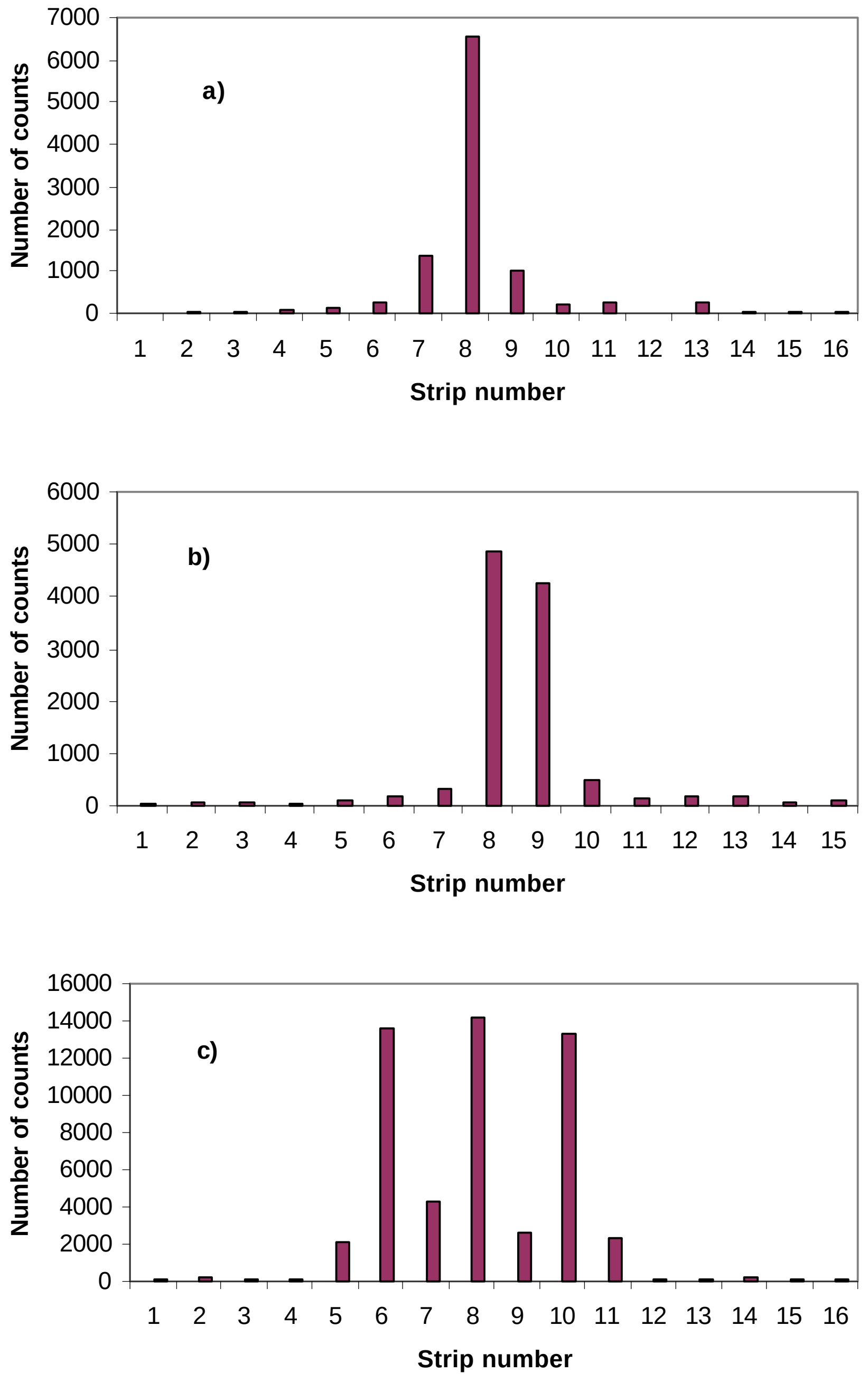

Fig. 7. Number of counts from various strips in the case when X-rays penetrated to the RPC through a 30 μm slit with the beam oriented along the strip #8 (a) and in between the strips #8 and #9 (b). The picture was periodically repeated when the beam was further moved perpendicularly to the strips. c) Number of counts from various strips for X-ray imaging of a 7 lp/mm phantom.